\documentclass[prd,a4paper,
showpacs,
twocolumn
]{revtex4}
\usepackage{graphicx}

\def\hbar{\hspace{0pt}\raisebox{1pt}{$-$} \hspace{-7pt} h}

\def\5{\overline 5}

\newcommand{\ba}{\begin{eqnarray}}
\newcommand{\ea}{\end{eqnarray}}
\newcommand{\no}{\nonumber}
\newcommand{\be}{\begin{equation}}
\newcommand{\ee}{\end{equation}}
\newcommand{\bea}{\begin{eqnarray}}
\newcommand{\eea}{\end{eqnarray}}

\begin{document}
\title{The Conformal Window of deformed CFT's in the planar limit}
\date{\today
}
\author{Luca~Vecchi\footnote{vecchi@lanl.gov}}
\affiliation{
Theoretical Division T-2, Los Alamos National Laboratory\\
  Los Alamos, NM 87545, USA}
\begin{abstract}
We discuss in the planar approximation the effect of double-trace deformations on CFT's. We show that this large class of models posses a conformal window describing a non-trivial flow between two fixed points of the renormalization group, and reveal the presence of a resonance which we associate to the remnant of a dilaton pole. As the conformal window shrinks to zero measure the theory undergoes a conformal phase transition separating a symmetric from a nonsymmetric phase. The recently conjectured strongly coupled branch of non-supersymmetric, non-abelian gauge theories with a large number of flavors is analyzed in light of these results, and a model for the strong branch is proposed. Some phenomenological implications in the context of unparticle physics are also emphasized.
\noindent
\end{abstract}
\preprint{LA-UR 10-01894}
\maketitle
%

\section{Introduction}

Conformal symmetry is a powerful tool for the particle physicist. On a genuinely theoretical level, the role of conformal symmetry is crucial when dealing with the quantum behavior of a particle theory. In fact, our understanding of quantum field theories is mostly entirely based on the simple scaling laws that these systems exhibit in the vicinity of fixed points of the renormalization group. This is at the heart of ordinary perturbation theory, which allows us to study the quantum properties of a system sufficiently close to its Gaussian fixed point. A more ambitious aim would be the analysis of deformations of non-trivial conformal field theories, which would eventually lead to a deeper understanding of the strongly coupled regime of a field theory. 

On a more phenomenological level, the importance of the conformal symmetry in formulating realistic models of dynamical electro-weak symmetry breaking was emphasized time ago by Holdom~\cite{Holdom}. These considerations found a concrete realization in the \textit{walking technicolor} paradigm, which is one of the most attractive scenarios for physics beyond the standard model.

In this paper we wish to address both phenomenological and theoretical issues in a simple, tractable framework.

The prototype of any 4-dimensional conformal field theory (CFT) is ${\cal N}=4$ SYM. Because of its large amount of supersymmetry, it is hard to imagine a direct application in the context of particle physics: non-supersymmetric CFT's would be much more desirable. Attempts in this direction have been recently guided by the gauge/gravity correspondence, and are typically based on the idea of deforming the bulk $AdS_5\times S^5$ geometry in such a way that the $AdS_5$ factor survives~\cite{String1}\cite{String2}. Orbifold projections of ${\cal N}=4$ SYM are explicit realizations of this program. In this case, one introduces an orbifold symmetry that projects out part of the original supersymmetric field content, thus (partially or totally) breaking supersymmetry. The resulting theory acquires the following structure:
\ba\label{theory}
{\cal L}_{CFT}+\frac{f}{2}{\cal O}^\dagger_{ij}{\cal O}^{ij}.
\ea
Here, ${\cal L}_{CFT}$ denotes the $SU(N)$ gauge theory directly inherited by ${\cal N}=4$ SYM, whereas ${\cal O}_{ij}$ is a gauge singlet scalar typically charged under some internal symmetries (the $ij$ indeces above). An analysis of these models reveals that ${\cal L}_{CFT}$ preserves the conformal symmetry of the original supersymmetric theory at leading order in $1/N$~\cite{Inheritance}, but the lack of supersymmetry implies the unavoidable emergence of counterterms of the form $f{\cal O}_{ij}^2$, which introduce a conformal anomaly already at leading order~\cite{Kleba}\cite{PR}. The resulting picture suggests that the $AdS_5$ factor of the dual string theory becomes generally unstable as soon as supersymmetry is broken~\cite{Instability1}\cite{Instability2}.

Similar conclusions generalize to any theory admitting a structure of the form~(\ref{theory}), with ${\cal L}_{CFT}$ being a (large $N$) CFT. Examples belonging to this class include -- in addition to the already mentioned orbifold projections of ${\cal N}=4$ SYM -- many known quantum field theories, such as the (massless) sigma model and the Gross-Neveu model~\cite{GrossNeveu}, and have interesting applications in particle physics. The Nambu-Jona Lasinio model for chiral symmetry breaking~\cite{NJL} represents perhaps the most popular example. A less celebrated application was proposed by Strassler~\cite{Strassler}, who showed that the theory~(\ref{theory}), in which ${\cal L}_{CFT}$ may be identified with ${\cal N}=4$ SYM, is one of a few known examples of non-supersymmetric models admitting naturally light scalars, and hence may represent an interesting laboratory for the model builder.

Our aim is to address the model-independent features encoded in the general structure~(\ref{theory}). In Section II we will discuss the beta function for $f$ at leading order in the planar approximation and analyze the phase structure hidden in~(\ref{theory}). We will see that the dynamics admits a flavor nonsymmetric and a flavor symmetric phase. The symmetric phase is particularly interesting, as it manifests two distinct behaviors depending on the strength of the renormalized coupling of the deformation. The first possibility is that the theory enters an IR conformal phase, in which case higher order terms in $1/N$ would be required in order to avoid triviality. The second possibility is that the dynamics enters a conformal window describing a non-trivial flow between two fixed points of the renormalization group. In Section III we will see that in this latter case the operator ${\cal O}_{ij}$ interpolates a resonant state with a mass controlled by the explicit CFT breaking parameter and the scaling dimension of ${\cal O}_{ij}$. We will identify this resonance as the remnant of the Nambu-Goldstone boson of dilatation invariance.

As the external parameters (number of flavors, colors, space-time dimension, etc.) vary, there exists a limit in which the conformal window shrinks to zero measure. In this limit the theory manifests a conformal phase transition (CPT)~\cite{Miransky}, separating the symmetric from the nonsymmetric realization. Conformality appears to be broken by \textit{classically} marginal deformations, and the theory naturally accounts for a hierarchical separation between UV and IR scales.

In Section IV we specialize on models with 4-fermion interactions. We show that these models manifest all of the general properties discussed above, and analyze in some details the gauged Nambu-Jona Lasinio model. We take advantage of these results to discuss the conformal window of 4-dimensional non-abelian gauge theories in Section V, and analyze the recently conjectured strongly coupled branch of these theories~\cite{CL}.

Our study has also immediate implications on the unparticle physics~\cite{Georgi}. In the last Section we propose a class of unparticle scenarios in which the effect of the Higgs vacuum expectation value on the CFT dynamics can be unambiguously computed in the planar limit. The resulting theory is characterized by a flow towards a new IR fixed point and a resonant behavior. 

We conclude with a brief summery of the main results.

\section{Double-trace deformations}

We are interested in studying the dynamics encoded in a theory of the type~(\ref{theory}), where ${\cal L}_{CFT}$ is a (large $N$) conformal field theory and ${\cal O}_{ij}$ a scalar operator of the CFT. We will be referring to ${\cal L}_{CFT}$ as the \textit{undeformed} theory, and to $f{\cal O}^\dagger_{ij}{\cal O}^{ij}$ as the \textit{deformation}.

In order for our analysis to be more than qualitative we make the following simplifying assumptions:
\begin{itemize}
\item [i)] the dynamics of ${\cal L}_{CFT}$ is controlled by a large $N$ expansion suppressing higher order correlators for the operator ${\cal O}_{ij}$;
\item [ii)] the couplings of the undeformed theory, collectively called $\lambda$, do not receive corrections from $f$ at leading order in the planar limit.
\end{itemize}
In the above, $N$ refers to either a gauge or a flavor symmetry, or both, whereas the couplings $\lambda$ are understood to be the $O(1)$ 't Hooft couplings of ${\cal L}_{CFT}$. For definiteness we can imagine that $\langle{\cal O}_{ij}{\cal O}_{kl}\rangle_{CFT}=O(N)$, in which case 
\ba\label{CC}
f=O\left(\frac{1}{N}\right)
\ea
for the theory to admit a sensible $N\gg1$ limit.

Both assumptions are satisfied in orbifold projections of ${\cal N}=4$ SYM, in the (massless) linear sigma model, and in the four-fermion models that we will analyze in detail in Section IV. The underlying reason is that all of these theories have undeformed lagrangians which can be written as single-trace operators of the form $Tr\left(\lambda\varphi^2+\lambda\varphi^4+\dots\right)$, where $\varphi$ is a generic field charged under an $U(N)$ symmetry and ${\cal O}_{ij}$ can be identified with a single-trace, $U(N)$ singlet field $Tr(\varphi_i\varphi_j)$. For instance, $\varphi_i$ denotes a scalar in the orbifolded ${\cal N}=4$ SYM and in the sigma model, while a fermion in the models discussed in Sections IV and V. One can then see that the \textit{double-trace deformation} does not induce any correction to the \textit{single-trace} action at leading order, i.e. the deformation $f{\cal O}_{ij}^2$ does not renormalize $\lambda$ in the planar limit. 

Under the assumptions i) and ii), the leading $1/N$ RG evolution of the deformed theory is completely determined by the running of the coupling $f$ of the double-trace deformation.

\subsection{The RG flow}

The simple set up illustrated above allows us to compute the beta function of the dimensionless coupling $\bar f=f\Lambda^{2\Delta-d}$ at leading order in $1/N$. This beta function first appeared in~\cite{PR} for $d=4$ and ${\cal O}_{ij}=Tr(\varphi_i\varphi_j)$ with $\varphi$ a scalar, where a detailed study of the phases of orbifold projections of ${\cal N}=4$ SYM was also presented. Instead of repeating the derivation of~\cite{PR} we employ a Wilsonian approach. We work in a generic space-time dimension $d$, and denote with $\Delta$ the scaling dimension of the operator ${\cal O}$ in the absence of the deformation, namely for $f(\Lambda)=0$. For simplicity we suppress the flavor indeces $i,j$.

We follow~\cite{Witten} and observe that in the evaluation of the effective action of the theory~(\ref{theory}) at the scale $\Lambda'<\Lambda$ one encounters the expansion
\ba\label{1}
1-\frac{f}{2}\int{\cal O}^2+\frac{f^2}{2}\int\int{\cal O}(x){\cal O}(y)\langle{\cal O}(x){\cal O}(y)\rangle+\dots.
\ea
At coincident points, $|x-y|\rightarrow0$, the $2$-point function induces a dependence on the RG scale, and thus renormalizes the coupling $f$. Higher order terms in the expansion~(\ref{1}) contribute similarly, but thanks to the factorization property of the $n$-point functions in the planar limit (this follows from our assumption i)),~(\ref{1}) constitutes the full, $O(1/N)$ contribution to the running of $f$. By normalizing the $2$-point function as
\ba\label{2point}
\langle{\cal O}(x){\cal O}(y)\rangle\equiv\frac{v}{|x-y|^{2\Delta}}\frac{\Gamma(d/2)}{2\pi^{d/2}},
\ea
with $v>0$ by unitarity, and integrating in the shell $1/\Lambda<|x-y|<1/\Lambda'$, it then follows that the coupling $f$ gets renormalized by the relation 
\ba\label{pass}
f(\Lambda)-\frac{v}{2\Delta-d}f^2(\Lambda)(\Lambda^{2\Delta-d}-\Lambda'^{2\Delta-d})=f(\Lambda'), 
\ea
plus subleading corrections in $1/N$. By differentiating with respect to $\Lambda'$ and taking the limit $\Lambda'\rightarrow\Lambda$ we obtain the beta function at leading order in $1/N$
\ba\label{beta1}
\Lambda\frac{d\bar f}{d\Lambda}=v\bar f^2+(2\Delta-d)\bar f.
\ea
Recalling that $v=O(N)$, we recover the consistency condition~(\ref{CC}).

In a similar way one can see that the coupling $f$ renormalizes the quadratic part in ${\cal O}$ hidden in the undeformed action~\footnote{We can think of the undeformed action as a quadratic theory in ${\cal O}$ and proceed as above.}. The new divergence can be reabsorbed into a redefinition of the operator, and leads to the emergence of an $O(1)$ anomalous dimension $\gamma_f=v\bar f$. The total dimension of the operator ${\cal O}$ is thus found to be
\ba\label{Dim}
\Delta_{\cal O}=\Delta+v\bar f.
\ea
We emphasize that~(\ref{beta1}) and~(\ref{Dim}) have been derived at leading order in $1/N$, but are valid at all orders in $\lambda,Nf$.

Before exploring the phase structure of~(\ref{theory}) we would like to stress that the above results can be straightforwardly generalized to a larger class of models in which the undeformed theory is not an actual CFT, provided the leading $2$-point function for the operator ${\cal O}$ is of the form~(\ref{2point}). This is precisely what happens in orbifold deformations of ${\cal N}=4$ SYM at leading order in $1/N$, for example, in which the undeformed theory cannot be regarded as a well defined field theory because of the necessity of the counterterm $f{\cal O}^2$. The above RG study should be modified to account for the possibility that the undeformed, single-trace theory generates terms $\propto(\Lambda^{d-2\Delta}-\Lambda'^{d-2\Delta}){\cal O}^2$ in the effective theory, where the factor is generally a function of $\lambda$. By introducing this contribution in~(\ref{pass}), the beta function~(\ref{beta1}) generalizes to
\ba\label{Beta}
\Lambda\frac{d\bar f}{d\Lambda}=v\bar f^2+(2\Delta-d)\bar f+a,
\ea
where $a,v,\Delta$ are functions of $\lambda$ and can be regarded as constants under our approximations. In particular, the anomalous dimension of the single-trace operator ${\cal O}$ already contains $\lambda$ corrections in~(\ref{2point}), and the total dimension in the presence of the deformation is again given by~(\ref{Dim}).

Equations~(\ref{Beta}) and~(\ref{Dim}) agree with the 1-loop results of~\cite{Kleba} and with the planar analysis of~\cite{PR}. 
In a subsequent publication we will rederive these results using the gauge/gravity correspondence~\cite{LV}.

\subsection{Phases of the deformed theory}

An inspection of~(\ref{beta1}) reveals that the theory~(\ref{theory}) admits two fixed points, an IR and an UV fixed point. Furthermore, from~(\ref{Dim}) we see that the dimension of the single-trace operator is $\Delta_{\cal O}=\Delta$ at the trivial FP, while $\Delta_{\cal O}=d-\Delta$ at the non-trivial FP. 

\begin{figure}
\begin{center}
\includegraphics[width=3.0in]{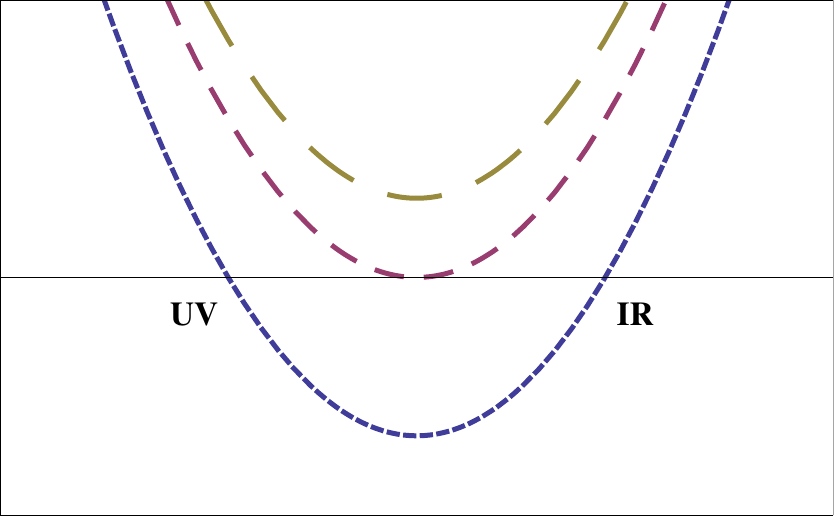}
\caption{\small  Schematic plot of the beta function~(\ref{Beta}) as a function of the coupling $\bar f$ for three non-trivial values of $a$. As $a$ increases (lower line to upper line) the (UV and IR) fixed points merge and then disappear. \label{fig1}}
\end{center}
\end{figure}

As soon as we allow a non-zero $a$, the picture can drastically change. The effect of a non-trivial $a$ has been studied in some details in~\cite{Kleba}\cite{PR}, and more recently in~\cite{CL}, and it is schematically shown in FIG.~\ref{fig1}. Let us analyze this generic situation.

If $a\neq0$ the beta function~(\ref{Beta}) admits real zeros only if $D=(\Delta-d/2)^2-va\geq0$. In this latter case we find two fixed points given by $v{\bar f}_\pm=d/2-\Delta\pm \sqrt{D}$. By evaluating the first derivative of the beta function at the fixed points we obtain the critical exponents
\ba\label{critexp}
\beta'_{\bar f}|_\pm=\pm2\sqrt{D},
\ea
and identify $\bar f_-$ as an UV attractive and $\bar f_+$ as an IR attractive fixed point. At $\bar f=\bar f_\pm$ the dimension of the operator ${\cal O}$ is
\ba\label{Dim'}
\Delta_\pm=\frac{d}{2}\pm \sqrt{D},
\ea
and satisfies the sum rule $\Delta_-+\Delta_+=d$. Notice that in order for the flow to be consistent with the unitarity bound on the dimension of the scalar operator, namely $\Delta\geq\frac{d-2}{2}$, we should further require that $0\leq\sqrt{ D}\leq 1$.

The renormalized coupling reads
\ba\label{fR}
\bar f(\Lambda)-\bar f_-=\frac{\bar f_+-\bar f_-}{1+\left(\frac{\Lambda}{\Lambda_0}\right)^{2\sqrt{D}}},
\ea
where generally $\bar f_+\geq\bar f_-$. For $D>0$ we thus identify two second-order phase transitions as a function of the \textit{temperature} $\bar f$. If $\bar f_-<\bar f<\bar f_+$ and $0<D<1$ the theory flows between the two fixed points and possesses a well defined continuum limit $\Lambda\rightarrow\infty$. The flavor symmetry carried by ${\cal O}$ is unbroken, and the theory is said to be in a conformal window. For $\bar f<\bar f_-$ or $\bar f>\bar f_+$ the coupling presents an IR or an UV Landau pole, respectively, at the scale
\ba\label{Landau}
\Lambda_L^2=\Lambda^2 \left|\frac{\bar f-\bar f_-}{\bar f-\bar f_+}\right| ^{\frac{1}{\sqrt{D}}}.
\ea
The phase $\bar f>\bar f_+$ can be regarded as an IR free theory with coupling $\bar f-\bar f_+$. In this case the theory is trivially symmetric and subleading corrections in $1/N$ are required in order to formulate a sensible continuum limit. For $\bar f<\bar f_-$ the theory develops a tachyonic instability and breaks the flavor symmetry. The actual vacuum of this phase is a model-dependent issue and cannot be addressed here.

If $D<0$ the beta function~(\ref{Beta}) has no real zeros. If we define $\Delta_I=\sqrt{|D|}$, the renormalized coupling can be written as
\ba\label{e}
v \bar f(\Lambda)=\left(\frac{d}{2}-\Delta\right)+\Delta_I\tan\left(\Delta_I\log\left(\frac{\Lambda}{\Lambda_0}\right)\right).
\ea
Eq.~(\ref{e}) tells us that now $\bar f$ interpolates between two Landau poles: the theory should be formulated in the finite energy range
\ba\label{Landau}
\Lambda_0e^{-\frac{\pi}{2\Delta_I}}<\Lambda<\Lambda_0e^{+\frac{\pi}{2\Delta_I}}.
\ea
In this case we expect the flavor symmetry to break down.

Formally, we can interpret the $D<0$ regime as describing a flow between two complex fixed points of the renormalization group~\cite{Kleba}\cite{PR}. Insisting on this interpretation it follows that the dimension of the single-trace operator at the complex fixed points is given by 
\ba\label{complex}
\Delta_\pm=\frac{d}{2}\pm i\Delta_I. 
\ea
In Section V we will elaborate on the connection between the Landau poles~(\ref{Landau}) and the breaking of conformality~(\ref{complex}) from a more general perspective. 

The relation between $\Delta_I$ and the violation of the Breitenlohner-Freedman bound~\cite{BF} in the dual bulk geometry has been analyzed in~\cite{Kleba}\cite{CL'}\cite{PR}\cite{CL}\cite{LV}.

Viewed as a function of $D$, and ultimately of the external parameters ($N_f,N,d,\dots$), the model~(\ref{theory}) manifests an asymptotically conformal symmetry for $D>0$, whereas conformal symmetry is completely lost for $D<0$. At $D=0$ the fixed points merge at the critical point $\bar f_-=\bar f_+=\bar f_c$ and the theory undergoes a conformal phase transition (CPT)~\cite{Miransky}, see the middle curve in FIG.~\ref{fig1}. In this limit the theory posses an IR-free realization for $\bar f-\bar f_c>0$ and an asymptotically safe realization for $\bar f<\bar f_c$. In the former case the theory is symmetric, whereas in the latter case the model develops an instability at the IR Landau pole and decays into a nonsymmetric vacuum. For $D=0$ the order parameter $\rho$ is then expected to be of the form
\ba\label{rho}
\rho&=&\theta(\bar f_c-\bar f)\Lambda\, e^{-\frac{1}{v(\bar f_c-\bar f)}}\\\no
&=&\theta(\Delta_c-\Delta_{\cal O})\Lambda\, e^{-\frac{1}{\Delta_c-\Delta_{\cal O}}},
\ea
where the scale in the nonsymmetric phase has been taken to coincide with the IR Landau pole for definiteness, and $\Delta_c=d/2$. Eq.~(\ref{rho}) presents an essential singularity at $\Delta_{\cal O}=\Delta_c$, a defining property of the CPT. 

For $D=0$ the model develops a natural separation between the UV and the IR scales, in analogy with QCD. Technically, this improvement in the degree of naturalness is due to the nature of the deformations driving the theory away from the critical point $\bar f_c$. A measure of the relevance of the deformation is given by the critical exponent~(\ref{critexp}) at the UV fixed point, which in the limit under study vanishes. Stated differently, for $D=0$ the double-trace deformation is classically marginal but quantum mechanically relevant, very much like the glueball field in an asymptotically free theory.

\section{Resonant behavior}

We can summarize the results obtained so far by writing down an expression for the 2-point function $G_R$ of the renormalized operator ${\cal O}_R$. The $2$-point function satisfies the Callan-Symanzik equation (where $q^2=-p^2$ denotes the Euclidean momenta)
\ba
\left(q\frac{\partial}{\partial q}-\beta_{\bar f}\frac{\partial}{\partial\bar f}\right)G_R=(2\Delta_{\cal O}-d)G_R.
\ea
Using the RG equation for the coupling~(\ref{Beta}) and the quantum dimension of the operator ${\cal O}$ given by~(\ref{Dim}), and focusing on the case $D>0$, we find 
\ba\label{GR}
&&G_R(p^2)=\\\no
&&\frac{i}{(\bar f_+-\bar f)(\bar f-\bar f_-)}\frac{(-p^2)^{\Delta-d/2}}{\left(\frac{\bar f-\bar f_-}{\bar f_+-\bar f}\right)^{\frac{\Delta-d/2}{\sqrt{D}}}+\left(-\frac{p^2}{\mu^2}\right)^{\Delta-d/2
}}.
\ea

The two-point function derived above manifests a resonant behavior at the scale
\ba
\left(-\frac{p_\sigma^2}{\mu^2}\right)^{\Delta-d/2}=-\left(\frac{\bar f-\bar f_-}{\bar f_+-\bar f}\right)^{\frac{\Delta-d/2}{\sqrt{D}}}.
\ea
The explicit solution reveals the presence of an infinite number of zeros
\ba\label{mass}
p_\sigma^2=\mu^2\left|\frac{\bar f-\bar f_-}{\bar f_+-\bar f}\right|^{1/\sqrt{D}}e^{i\pi\left(\frac{1+2k}{\sqrt{D}}+1\right)},
\ea
(here $k=0,\pm1,\pm2,\dots$) which can be shown to be RG invariant using~(\ref{Beta}). Outside the range $\bar f_-<\bar f<\bar f_+$ the poles are tachyonic, as anticipated by the RG analysis. Let us focus on the physics of the conformal window, i.e. the case $\bar f_-<\bar f<\bar f_+$.

The zeros~(\ref{mass}) are found outside the physical sheet (we are tacitly assuming that $0<\sqrt{D}<1$), as expected for resonant poles; they lie on a circle and are separated by an angular distance $\pi/\sqrt{D}>\pi$. This latter observation ensures that, in practice, only a state can be phenomenologically relevant. The relevance of the resonance can be measured by the smallness of the phase in the exponential in~(\ref{mass}), that is to say the vicinity of the pole to the real axis. This is tantamount to saying that the ratio $\Gamma_\sigma/m_\sigma$ -- we defined the mass and width of the resonance as $p_\sigma=m_\sigma-i\Gamma_\sigma/2$ -- is small. One then immediately sees that
\ba
\frac{\Gamma_\sigma}{m_\sigma}=-2\tan\left(\frac{1+2k+\sqrt{D}}{2\sqrt{D}}\pi\right).
\ea
A small phase can only be achieved for $\sqrt{D}\rightarrow1$ and if $k=-1$. In the latter case
\ba
\frac{\Gamma_\sigma}{m_\sigma}\simeq\frac{1-\sqrt{D}}{\sqrt{D}}\pi,
\ea
which is positive, indicating that the resonance lies in the second Riemann sheet, below the cut, as general quantum field theory arguments require. The physical interpretation of the result is that as $\sqrt{D}\rightarrow1$ from below the operator ${\cal O}$ interpolates a scalar field with decreasing width, in agreement with the fact that in this limit the UV dimension of the operator approaches that of a free scalar field.

Because in the broken phase $\bar f<\bar f_-$ the longitudinal excitation of ${\cal O}$ plays the role of the would-be Nambu-Goldstone boson of conformal invariance, we may refer to the resonance found in the conformal window as a remnant of the dilaton pole.

The resonance has a mass $O(1)$ and is generally broad, as opposed to ordinary mesons which have widths suppressed by powers of $1/N$ and arise in confining theories. It is interesting to discuss the physics leading to the appearance of the resonance from a broader perspective. Let us denote by $I$ the 1PI effective action of the CFT. By focusing on the dynamics of the operator ${\cal O}$ and recalling that at leading order in $1/N$ the only surviving correlators are the two-point functions, we can think of the effective action as a quadratic functional of the classical field $\alpha=\langle{\cal O}\rangle$. The generating functional reads
\ba
I[\alpha]=\frac{1}{2}\int \alpha\frac{1}{G_\pm}\alpha,
\ea
where the Euclidean propagator is $G_-=+q^{2\Delta-d}$ if $\Delta<\frac{d}{2}$, and $G_+=-q^{2\Delta-d}$ if $\Delta>\frac{d}{2}$ (the different sign is a consequence of~(\ref{2point}) and positivity of $v$). We now include an arbitrary function $W(\alpha)$ and define a deformed action
\ba\label{eff}
I_f[\alpha]=I[\alpha]+\int W(\alpha),
\ea
where $f$ is a collective name for the parameters entering $W$. In the cases considered here the deformation is given by $W=f\alpha^2/2$, and acts as an effective mass term for the field $\alpha$. The deformed propagator is easily obtained (in Euclidean space) as
\ba\label{Dyson}
G=\left(\frac{1}{G_\pm}+f\right)^{-1}=\frac{G_\pm}{1+fG_\pm},
\ea
which is clearly of the form~(\ref{GR}). Depending on the sign of $f$ and the scaling dimension $\Delta$ the mass is tachyonic or not. Precisely, for $\Delta<\frac{d}{2}$ stability requires $f>0$, while for $\Delta>\frac{d}{2}$ it requires $f<0$.

\section{The four-fermion interaction}

It is instructive to study in some details explicit field theories that present all the general properties described in the previous Section. Here we specialize on the theory of 4-fermion contact interactions. This study will turn out to be useful when discussing the conformal window of non-abelian gauge theories in $d=4$.

\subsection{The NJL model in $2<d<4$}

Consider the Nambu-Jona Lasinio (NJL) model 
\ba\label{GN}
{\cal L}=\bar\psi i\partial\psi+\frac{f}{2}(\bar\psi\psi)^2,
\ea
where the fermionic field is charged under a \textit{color} $U(N)$ global symmetry. The theory is evidently of the form~(\ref{theory}), where the undeformed action is trivial and ${\cal O}$ can be identified with the $U(N)$ singlet $\bar\psi\psi$. In the absence of a mass term, the above model also admits a discrete chiral symmetry under which $\bar\psi\psi\rightarrow-\bar\psi\psi$. It is straightforward to generalize the discussion to larger chiral groups. For $d=2$ the above theory is the Gross-Neveu model~\cite{GrossNeveu}.

The engineering dimensions of the field and coupling are $[\psi]=\frac{d-1}{2}$ and $[f]=2-d$, and the coupling is classically irrelevant for $d>2$. We will see that the beta function for the dimensionless coupling $\bar f=f\mu^{d-2}$ receives a negative contribution from quantum fluctuations. As a result, the above model would develop, in addition to the Gaussian fixed point, a non-trivial UV fixed point. At the UV fixed point the 4-fermion operator becomes relevant, and the theory admits a well defined continuum limit.

The action~(\ref{GN}) is not renormalizable as a perturbative expansion in $f$. Nevertheless it is found to be renormalizable as an expansion in the dimensionless parameter $1/N$ for $d<4$~\cite{Parisi},\cite{Gross} (see also~\cite{Rosenstein} and~\cite{Kogut}). The inequality $d<4$ follows from power counting renormalizability in~\cite{Parisi}, while as a consequence of unitarity from our previous results. Indeed, the sum rule $\Delta_++\Delta_-=d$, which we found to be a generic property of these models at leading order in $1/N$, requires that the UV dimension of the operator ${\cal O}$ -- assuming an UV fixed point exists -- is $\Delta_-=d-\Delta_+=1$, with $\Delta_+=d-1$ being the dimension at the Gaussian fixed IR point. The unitary bound $\Delta_-\geq\frac{d-2}{2}$ thus translates into a constraint on the dimensionality of the space-time, as anticipated. In the following we will analyze in some details the model for dimensions $2<d<4$.

As a preliminary step we evaluate the 2-point function $G_+(p)$ for the operator ${\cal O}\equiv\bar\psi(x)\psi(x)$ at the Gaussian fixed point. The diagram contributing to $G_+$ corresponds to the one-bubble graph, and will play an essential role in what follows. We find
\ba\label{Gross-Neveu}
G_+(p)&=&(-1)\int\frac{d^dk}{(2\pi)^2}\textit{Tr}\frac{i}{\not k}\frac{i}{\not k-\not p}\\\no
&=&i\frac{1}{\bar f_*}\left(-p^2\right)^{d/2-1},
\ea
with $\mu$ being the RG scale. The integral has been evaluated using dimensional regularization and we defined the quantity $\bar f_*$ as
\ba
\frac{1}{\bar f_*}=\frac{N}{(4\pi)^{d/2}}\frac{\Gamma\left(\frac{d}{2}\right)}{\Gamma(d)}\pi d(d-1)\left(\frac{-1}{\sin(d\pi/2)}\right).
\ea
In the range $2<d<4$ we have $\sin(d\pi/2)<0$ and consequently $\bar f_*>0$.

As a consistency check of~(\ref{Gross-Neveu}) we notice that $G_+$ trivially satisfies the Callan-Symanzik equation 
\ba
q\frac{\partial}{\partial q}G_+=(2\Delta_+-d)G_+,
\ea
where $q^2=-p^2$ is the Euclidean momenta. As a non-trivial check one can show that the correlator is compatible with unitarity. Because the correlator $G_+$ enters in physical amplitudes, the optical theorem requires the imaginary part of $i G_+$ to be positive when evaluated on the physical region $p^0\geq0$ by analytic continuation from the upper half plane~\cite{GI}. In our specific example~(\ref{Gross-Neveu}) we obtain
\ba
\textit{Im}\left[iG_+\right]=\frac{N}{(4\pi)^{d/2}}\frac{\Gamma\left(\frac{d}{2}\right)}{\Gamma(d)}\pi d(d-1)|p^2|^{d/2-1},
\ea
which is manifestly positive for any dimension $d>1$.

When considering the interacting theory, a crucial observation to be made is that at leading order in the $1/N$ expansion the relevant diagrams to be summed correspond to 1-bubble diagrams, i.e. the calculation is performed in the ladder approximation, the other graphs are either subleading or they vanish because of the traceless nature of the Dirac matrices. The one-bubble diagram determines all the quantities of our interest at the leading $1/N$, in particular the amplitude ${\cal A}(p^2)$ for the scattering $2\psi\rightarrow2\psi$ and the connected propagator $G(p^2)$. The explicit expressions for these quantities are found to be
\ba\label{A}
i{\cal A}(p^2)=\frac{if}{1-ifG_+},
\ea
and
\ba\label{G+}
G(p^2)=\frac{G_+}{1-ifG_+}.
\ea
In the above expressions and in the following discussion we will omit for simplicity the spinor structure. Notice that the leading $O(f)$ term in the amplitude agrees with the classical expectations; yet the result is non-perturbative in the coupling $f$. The 2-point function has the anticipated form~(\ref{Dyson}).

The above expressions~(\ref{A}) and~(\ref{G+}) are written in terms of the bare quantities. In order to appreciate the quantum content of the results we will need to express the correlators in terms of the renormalized operator ${\cal O}_R$. The renormalization procedure is carried out by regularizing the integrals using dimensional regularization and imposing physical renormalization conditions at some RG scale $\mu$. As the operator $\psi$ is not renormalized at leading order in $1/N$ (that is to say, the undeformed action in~(\ref{theory}) is not affected by the deformation at leading order), it will suffice to consider two renormalization conditions. The latter are chosen as follows
\ba\label{RC}
\bar f(\mu)={\cal A}(-\mu^2)\mu^{d-2}\quad\quad\langle{\cal O}_R\psi\bar\psi\rangle |_\mu=1.
\ea
Their physical interpretation is straightforward. The first condition defines the renormalized and dimensionless coupling $\bar f$ at the scale $\mu$ as the strength of the 4-fermions interaction at the Euclidean scale $p^2=-\mu^2$. The second condition determines the wavefunction renormalization of the operator ${\cal O}$.

Now, using the explicit expressions~(\ref{A})(\ref{G+}) and relating the bare quantities to the renormalized ones via the relations
\ba
f=\mu^{2-d}Z_f\bar f\quad\quad{\cal O}=Z_{\cal O}{\cal O}_R,
\ea
we find that~(\ref{RC}) require
\ba
Z_f^{-1}=Z_{\cal O}=1-\frac{\bar f}{\bar f_*}.
\ea
From the independence of the bare coupling and the bare operator on the renormalization scale we derive the RG equation for the renormalized coupling and the anomalous dimension of the renormalized operator:
\ba\label{betaGN}
\mu\frac{d\bar f}{d\mu}&=&(d-2)\bar f\left(1-\frac{\bar f}{\bar f_*}\right)\\\no
\mu\frac{d}{d\mu}\log Z_{\cal O}&=&(2-d)\frac{\bar f}{\bar f_*}.
\ea
These are found to be particular cases of the more general results~(\ref{beta1}) and~(\ref{Dim}), with $\bar f_*$ being an UV fixed point. In the limit $d=2+\epsilon$, $\epsilon\ll1$, one finds $\bar f_*=\pi\epsilon/N$ in agreement with a perturbative expansion in $\epsilon$ (see for example~\cite{Zinn}).

The renormalized correlator is~(\ref{GR})
\ba
G_R(p^2)=\frac{i}{\bar f_*-\bar f}\frac{(-p^2)^{d/2-1}}{1-\frac{\bar f}{\bar f_*}+\frac{\bar f}{\bar f_*}\left(-\frac{p^2}{\mu^2}\right)^{d/2-1}},
\ea
and the quantum dimension of the operator, 
\ba
\Delta_{\cal O}=d-1+(2-d)\frac{\bar f}{\bar f_*},
\ea
satisfies the sum rule discussed after~(\ref{Dim'}). The propagator, as well as the amplitude ${\cal A}$ manifest the resonant behavior illustrated in the previous Section. 

By evaluating the critical exponents~(\ref{critexp}) we find $2\sqrt{D}=d-2$, and infer that the NJL model naturally accounts for a separation between the IR scale $f$ and the UV cutoff $\Lambda$ only for $d=2$. For $d<4$ the 4-fermion interaction has a coupling $f$ naturally at the cutoff scale. In order to formulate a non-trivial theory for the NJL model in $d<4$, an amount of tuning is required. This is analogous to the naturalness problem of the linear sigma model, the only difference is that the unnatural deformation in the present case is irrelevant at the fixed point $\bar f=0$ (but relevant at the UV fixed point!). This fine-tuning is at the origin of the naturalness problem plaguing the top-condensate models for electro-weak symmetry breaking~\cite{topQuark1}\cite{topQuark2}.

\subsection{Quenched QED in $d=4$}

We now move to $d=4$ and consider massless QED in the quenched, planar approximation. This theory was first proposed by W. A. Bardeen, C. N. Leung, and S. T. Love as a toy model for spontaneous conformal \textit{and} chiral symmetry breaking~\cite{Bardeen1}. A summery of their findings can be found in~\cite{Bardeen}. As we will review below, a consistent description of the model requires an explicit, rather than spontaneous breaking of the conformal symmetry, and no massless dilaton emerges. Nevertheless, the model is appealing for at least two important reasons. First, it represents an interesting laboratory for the study of spontaneous chiral symmetry breaking in non-abelian gauge theories with slowly varying (walking) coupling; second, it is an explicit 4-dimensional realization of the framework discussed in Section II. Because the literature on the subject is considerable, we refer the reader to a recent review by Koichi Yamawaki for further references~\cite{review}.

In the quenched approximation the fermion loops are suppressed and the 't Hooft coupling $\lambda=\frac{g^2}{16\pi^2}$ does not run at leading order:
\ba\label{betae}
\beta_\lambda=0.
\ea
To gain some insight into the non-perturbative dynamics of the theory we can analyze the homogeneous Schwinger-Dyson equation in Euclidean space in the rainbow approximation
\ba\label{SD}
\Sigma(q)=3g^2\int\frac{d^4k}{(2\pi)^4}\frac{1}{(q-k)^2}\frac{\Sigma(k)}{k^2+\Sigma^2(k)}.
\ea
In the above equation $\Sigma$ denotes the dynamical fermion mass: if the integral equation admits a solution with $\Sigma(0)\neq0$, then chiral symmetry spontaneously breaks down. Eq.~(\ref{SD}) can be reduced to a differential equation with appropriate boundary conditions, and eventually solved numerically. By defining
\ba
\Sigma(q)=e^tu(t+t_0),
\ea
where $t=\log q$ and $t_0=-\log\Sigma(0)$, we find that~(\ref{SD}) becomes
\ba
u''+4u'+3u+\frac{\lambda}{\lambda_c}\frac{u}{1+u^2}=0;\quad\quad \lambda_c=\frac{1}{12}.
\ea
For brevity we will not report the boundary conditions to be imposed on the solutions, the interested reader can find them in the above references or derive them by direct computation. Because $\Sigma\rightarrow0$ at large momenta, the solution for $\lambda<\lambda_c$ can be written as
\ba\label{Sigma}
\Sigma(q)&\sim&\frac{A}{q}\sinh\left(\sqrt{1-\frac{\lambda}{\lambda_c}}\log q+B\right)\\\no
&=&\frac{m}{q^\gamma}+\frac{\langle\bar\psi\psi\rangle}{q^{2-\gamma}},
\ea
with $\gamma=1-\sqrt{1-\frac{\lambda}{\lambda_c}}$. A similar form is found for $\lambda>\lambda_c$, but with the the hyperbolic sin replaced by sin. The first and second terms of~(\ref{Sigma}) refer to an explicit and spontaneous breaking of the chiral symmetry, respectively. It is the boundary conditions that fix the integration constants, and thus determine if spontaneous chiral symmetry breaking actually occurs. This important point will be analyzed in some details below. For the moment we notice that equation~(\ref{Sigma}) immediately tells us that the total dimension of ${\cal O}=\psi_R^\dagger\psi_L$ in the planar limit is
\ba\label{+}
\Delta_+=2+\sqrt{D};\quad\quad D\equiv1-\frac{\lambda}{\lambda_c}.
\ea
From this result we learn two important facts. First, since the theory is conformal in the leading approximation, the two-point function for ${\cal O}$ in quenched, planar QED is simply given by $G_+(q)=-q^{2\Delta_+-4}$. Second, at the leading order the operator ${\cal O}^\dagger{\cal O}$ is marginal at the critical point $\lambda=\lambda_c$.

The last observation led W. A. Bardeen et al. to the study of the \textit{gauged} NJL model:
\ba\label{gaugedNJL}
-\frac{1}{4g^2}F^2+\bar\psi iD\psi+\frac{f}{2}{\cal O}^\dagger{\cal O}.
\ea
A consistent way to deal with the four-fermion operator within the quenched approximation has been developed by these authors, and amounts to sum all the bubble graphs, in analogy with a large $N$ analysis. In the leading approximation the condition~(\ref{betae}) is preserved and the deformed two-point function in the presence of the bare coupling $f$ can be found as in the previous subsection:
\ba\label{qui}
G(q)&=&\frac{G_+}{1-fG_+}\\\no
&\sim&\left\{ \begin{array}{ccc}  -q^{2\Delta_+-4}\quad\quad q^{2\sqrt{D}}f\ll1\\
+q^{4-2\Delta_+}\quad\quad q^{2\sqrt{D}}f\gg1. \end{array}\right.
\ea
In the second line we extracted the leading non-analytic part in the Euclidean momentum for the two limiting regions (IR and UV, respectively). The asymptotic form of the Green's function reveals the scaling dimension of the composite operator ${\cal O}$. Consistently with our expectations we find that the dimension is $\Delta_+$ in the far IR, where the four-fermion operator is irrelevant, see~(\ref{+}), whereas it reduces to 
\ba
\Delta_-=2-\sqrt{D}<2
\ea
in the far UV~\cite{MY}. Notice that this result implies that the four-fermion operator is relevant in the UV. 

QED plus the four-fermion interaction describes in the planar limit a flow between two fixed points of the type described in Section II. Following the same procedure used for the NJL model we define a renormalized 4-fermion coupling $\bar f(\mu)$ as the (properly normalized) amplitude for the $2\psi\rightarrow2\psi$ scattering at the Euclidean scale $\mu$. The resulting beta function reads
\ba\label{fbar}
\mu\frac{d\bar f}{d\mu}=2\sqrt{D}\bar f\left(1-\frac{\bar f}{\bar f_*}\right).
\ea
This expression is analogous to the beta function of the NJL model, the only difference being found in the critical exponents~\cite{Kondo}.

We can analyze the phase structure of the model by studying the gap equation in the general case $f\neq0$. Defining a dimensionless coupling $\bar f'$ as
\ba
\bar f'=\frac{f}{2\pi^2}\Lambda^2,
\ea
where $\Lambda$ is an UV cutoff, one finds that the solutions of the Schwinger-Dyson equation can be implicitly written in the weakly coupled regime $\lambda<\lambda_c$ as~\cite{Bardeen}\cite{Takeuchi}
\ba\label{weak}
\left\{ \begin{array}{ccc}  \tanh\theta=\frac{\bar f'+\frac{\lambda}{\lambda_c}}{\bar f'-\frac{\lambda}{\lambda_c}}\sqrt{D} \\
\theta=\sqrt{D}\log\left(\frac{\Lambda e^\delta}{\Sigma(0)}\right), \end{array}\right.
\ea
and analogously for the strongly coupled regime $\lambda>\lambda_c$ if $\tanh$ is substituted by $\tan$ and $D$ by $|D|$.
In the above expression $\delta$ is an $O(1)$ function of $\lambda$, and can be regarded as a constant. We immediately see that, in the weak regime $\lambda<\lambda_c$, a consistent solution requires $f\neq0$ (if the physical condition $\Sigma(0)\ll\Lambda$ holds, $\tanh$ must be positive). No chiral symmetry breaking occurs in the absence of the four-fermion contact term. In the strong regime $\lambda>\lambda_c$ a solution can in principle be found even for $f=0$, but it requires a running of $\lambda$ (known as Miransky scaling) in order to ensure finiteness of $\Sigma(0)$ as $\Lambda\rightarrow\infty$. This latter behavior would be in contrast with our expectations from the planar approximation, and should be regarded as inconsistent. Chiral symmetry breaking in quenched QED can only occur if $f\neq0$. As we will see in the next Section, this conclusion does not generalize to the non-abelian case, where a running of the 't Hooft coupling is allowed already at leading order in the planar limit.

Differentiating~(\ref{weak}) with respect to the cutoff scale $\Lambda$ and requiring that $\Sigma(0)$ and $\lambda$ do not pendent on $\Lambda$, we find
\ba\label{betaQED}
\Lambda\frac{d\bar f'}{d\Lambda}&=&-\frac{1}{2}\left(\bar f'-\bar f'_-\right)\left(\bar f'-\bar f'_+\right),\\\no
\bar f'_\pm&=&\left(1\pm\sqrt{D}\right)^2.
\ea
The same equation may be found by differentiating the solution of the strongly coupled regime: the RG flow does not depend on the phase. The non-zero constant term in the beta function (of the order $\lambda^2$) seems to imply that a 4-fermion counterterm is required to perturbatively renormalize QED. This apparent inconsistency may be an artifact of the renormalization procedure leading to~(\ref{betaQED}), and it has no physical impact on our analysis. In particular, the coupling $\bar f'$ is in 1-to-1 correspondence with the renormalized coupling $\bar f$ introduced above, and therefore describes the very same physics. This can be immediately seen by observing that the critical exponents of~(\ref{betaQED}) and~(\ref{fbar}) at the two fixed points do coincide:
\ba
\beta'_{\bar f}=\pm2\sqrt{D}.
\ea
The phase structure is now evident, and in perfect agreement with our general analysis. 

The results presented in this Section suggest that the gauged NJL model in $d=4$ may admit a continuum limit $\Lambda\rightarrow\infty$ if defined at one of the UV fixed points $(\lambda=\lambda_c,\bar f'=1)$ or $(\lambda<\lambda_c,\bar f'=\bar f'_+)$, see~\cite{MY}. At each of these, the 4-fermion operator at leading order in the planar limit has UV dimension $2\Delta_-=2(2-\sqrt{D})\leq4$, and it is hence a (quantum mechanically) relevant operator. This observation strongly relies on the non-triviality of the undeformed theory.

\section{Non-abelian gauge theories}

We now wish to elaborate on the generalization of the results presented for quenched QED to the case of a non-abelian $SU(N)$ dynamics in the planar approximation. In particular we would like to speculate on the possibility of formulating a consistent continuum limit for the non-abelian version of~(\ref{gaugedNJL}).

The main difference from the abelian case evidently consists in the presence of a non-trivial beta function for the 't Hooft coupling $\lambda$ already at leading order in the $1/N$ expansion. If we insist on asymptotic freedom, the theory would be naturally formulated at the trivial UV fixed point $\lambda=0$ at which the 4-fermion interaction is clearly irrelevant. The arguments in favor of the renormalizability of the deformed theory would thus come into question (see however~\cite{KSY}\cite{Kugo}). A less conventional approach towards the formulation of a consistent theory for the contact 4-fermion interaction would be to give up asymptotic freedom. We will address this latter possibility in this Section.

\subsection{The conformal window}

It is widely believed that non-supersymmetric $SU(N)$ gauge theories in $d=4$ dimensions have the following qualitative zero temperature phase structure as a function of the ratio $x=N_f/N$, where $N_f$ denotes the number of massless flavors~\cite{ATW}\cite{Miransky}\cite{ARTW}\cite{Sannino}\cite{Poppitz}. For small $x$ the theory essentially behaves as pure Yang-Mills and presents both chiral symmetry breaking and confinement. This is the case of real QCD. As $x$ increases above a critical value $x_c$ the theory enters an IR conformal regime, where no chiral symmetry breaking nor confinement take place, but still manifests asymptotic freedom. In this case the theory is said to be in the conformal window. If the parameter $x$ further increases and exceeds an upper bound $x_{af}$ the theory looses asymptotic freedom and becomes somewhat ambiguous. In this case the theory is IR free.

To be as concrete as possible we briefly review the analysis of the two-loop beta function for the 't Hooft coupling
\ba
\lambda=\frac{g^2N}{16\pi^2}.
\ea
We work in the Veneziano limit ($N,N_f\gg1$ with $x$ fixed)~\cite{Veneziano}, in which case $x$ can be seen as a continuous variable. For fermions in the fundamental representation one finds
\ba\label{QCDcoupling}
\beta_\lambda=-b_1\lambda^2-b_2\lambda^3\quad\quad\left\{ \begin{array}{ccc}  b_1=\frac{2}{3}(11-2 x) \\\\
b_2=\frac{2}{3}(34-13 x). \end{array}\right.
\ea
We remind the reader that the two-loop result is scheme-independent, and that there exist a set of renormalization conditions in which eq.~(\ref{QCDcoupling}) is the exact perturbative expression of the beta function. 

An inspection of~(\ref{QCDcoupling}) shows that asymptotic freedom is preserved as long as $b_1>0$, which requires $x<x_{as}=\frac{11}{2}$. Furthermore, for $x\leq x^*=\frac{34}{13}<x_{as}$ we have $b_2\geq0$. Thus, there exists a range $x^*<x<x_{as}$ such that $b_1>0$ but $b_2<0$ in which the theory is asymptotically free but develops an IR fixed point
\ba\label{BZ}
\lambda_{IR}=-\frac{b_1}{b_2}.
\ea
For $x>x_{as}$ it follows that $b_1,b_2<0$: asymptotic freedom is lost, and the only zero we find is the Gaussian IR fixed point.

The two-loop analysis breaks down when non-perturbative effects come into play. This is expected to occur when, as $x$ decreases, the IR fixed point $\lambda_{IR}$ reaches a critical coupling $\lambda_c$ above which the chiral condensate forms~\cite{Appelquist}. The existence of a critical number of flavors $x_c$ at which $\lambda_{IR}=\lambda_c$ can be inferred by a study of the gap equation similar to the one performed for the $U(1)$ case, and the estimates of~\cite{ATW} suggest that it should be reached at $x=x_c>x^*$. As the chiral symmetry breaks down, a dynamical mass for the quarks is induced and the theory confines. In this case the two-loop prediction should not be trusted below $x_c$, and the lower end of the conformal window should be signaled by $x_c$ rather than by the perturbative estimate $x^*$.

A scheme-independent characterization of the conformal window may be given in terms of the IR scaling dimension $\Delta$ of the quark bilinear: 
\ba\label{range}
\Delta_c\leq\Delta<3.
\ea
Whereas the upper end can be estimated using perturbation theory ($\lambda_{IR}\equiv\lambda_{BZ}\ll1$ for $x=\frac{11}{2}-\epsilon$)~\cite{BZ} 
\ba
\Delta_{BZ}=3-3\lambda_{BZ}<3\quad\quad\lambda_{BZ}=\frac{4\epsilon}{75}\ll1,
\ea
there are indications (see for instance the QED model) that $\Delta_c=2$ in fact represents the non-perturbative condition for the lower end~\cite{CG}.

\subsection{The strongly coupled branch}

\begin{figure}
\begin{center}
\includegraphics[width=3.0in]{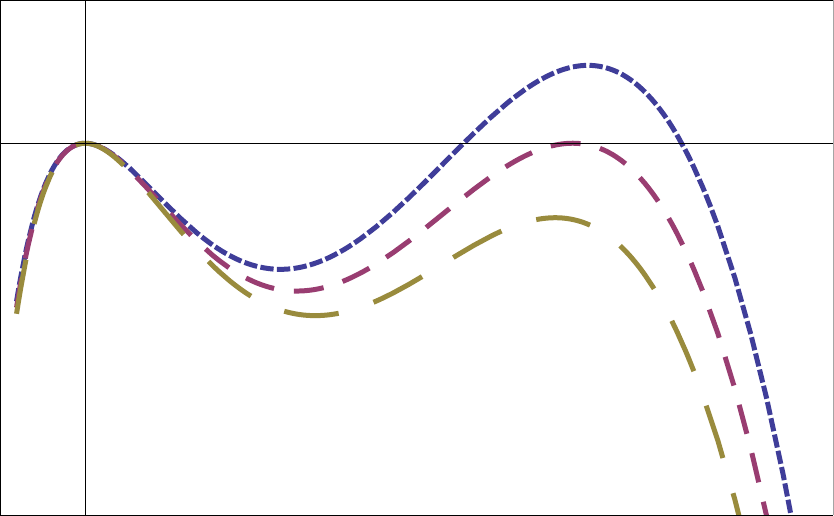}
\caption{\small  Conjectured beta function for a non-supersymmetric strong dynamics as a function of the coupling for three different values of $x$ in the vicinity of the critical value $x_c$. In the lower line $x<x_c$, in the middle line $x=x_c$, and in the upper line $x_c<x<x_{as}$. \label{fig2}}
\end{center}
\end{figure}

In~\cite{CL} it has been conjectured that the beta function $\beta_\lambda$ varies as a function of $x$ in a continuous way as illustrated in~FIG.~\ref{fig2} for $x\sim x_c$. A beta function with this qualitative behavior was analyzed in~\cite{Tuominem}. The picture reminds us of the phase diagram first proposed in~\cite{BZ} and then corrected by~\cite{Miransky} to account for the presence of the critical coupling $\lambda_c$. In these latter works, however, the non-trivial UV fixed point was understood as an artifact of the lattice analysis (bulk transition).

The picture~FIG.~\ref{fig2} may be seen as a consequence of the continuity of $\beta_\lambda$ as a function of $x$ and of the assumption that $\lambda_c$ is finite. In order to make an unambiguous statement, we should be able to formulate the conjecture in terms of physical quantities.

If the picture illustrated in~FIG.~\ref{fig2} is correct, QCD should admit a strongly coupled branch $\lambda>\lambda_{IR}$ in which a new, non-trivial UV fixed point $\lambda_{UV}$ would emerge. The phase transition between the conformal window and chiral symmetry breaking should then be signaled by a merger of two fixed points similar to the one studied in a previous Section as a function of $D$, the discriminant of~(\ref{Beta}). A physical characterization of the phase transition is therefore provided by the occurrence of a CPT (the Berezinskii-Kosterlitz-Thouless scaling of~\cite{CL}) separating a \textit{chirally-symmetric} conformal phase ($x>x_c$) from a \textit{chirally-nonsymmetric} confining phase ($x<x_c$).

As noted in~\cite{CL}, the physics is radically different in a supersymmetric realization. In our language this is a consequence of the fact that no CPT occurs in supersymmetric QCD, where the lower end of the conformal window separates a \textit{chirally-symmetric conformal} phase (at $x>\frac{3}{2}$) from a \textit{chirally-symmetric confining} phase (at $x<\frac{3}{2}$)~\cite{Seiberg}.

The occurrence of a CPT is intimately connected with a loss of conformality, and ultimately with the existence of a critical dimension $\Delta_c$ below which the chiral symmetry breaks down. We can formalize this statement as follows.

Let us assume that the picture emerging from~FIG.~\ref{fig2} is correct and denote with $\Delta(x)$ the IR dimension of the quark bilinear. We think of $\Delta$ as a continuous function of $x$. In the limit $x\rightarrow x_c^+$ the two (real) zeros of the beta function annihilate. Consistently, in the limit $x\rightarrow x_c^+$, $\Delta(x)$ should approach the critical dimension $\Delta_c=\Delta(x_c)$ at which conformality is lost. Now, by applying the same reasoning presented in Section II for the special case of a second order beta function, we can imagine that for $x<x_c$ the zeros of the beta function become (formally) complex quantities. The same conclusion would hold for the dimension $\Delta(x)$ at the \textit{formal} "IR fixed point", which in this case would read $\Delta=\Delta_R+i\Delta_I$, with $\Delta\rightarrow\Delta_c$ for $x\rightarrow x_c^-$. Insisting with this interpretation we are thus driven to consider the physical implications of a complex conformal weight. 

We expect that for $\Delta_I\rightarrow0$ the (Euclidean) 2-point function of the quark bilinear is approximately given by the CFT expression $G=Cq^{2\Delta-d}$, where $\Delta$ is now complex and $C$ is (generally a function of $\Delta$) consistent with unitarity when $\Delta_I=0$. With this notation, the optical theorem $Im[G]\geq0$ requires
\ba\label{phase}
Im\left[Ce^{i(\Delta_I\log|p^2|-\pi(\Delta_R-d/2))}\right]\geq0,
\ea
and can only be satisfied in a limited range of momenta. We can estimate this energy range by observing that under our assumptions the leading imaginary part of the correlator is given by the phase in~(\ref{phase}), the latter being always dominated by the $\log|p^2|$ term for sufficiently large/small momenta. Given a sign of $C$ compatible with unitarity when $\Delta_I=0$, the bound becomes simply $\sin(\Delta_I\log|p^2|)>0$ or $\sin(\Delta_I\log|p^2|)<0$ depending on the signs of $\Delta_I$ and $C$. If we insist in making sense of a theory with complex dimensions in a small neighborhood of $|\Delta_I|\ll1$, a conservative implication of unitarity is then $-\pi<\Delta_I\log|p^2|<\pi$: the theory must be defined between an IR and an UV cutoff, $\Lambda_{IR}$ and $\Lambda_{UV}$. These cutoffs are ambiguous quantities in the sense that their definition depends on the scale hidden in the $\log$. An RG invariant consequence of the unitarity bound can hence be expressed as a relation between the IR Landau pole and the UV cutoff:
\ba\label{rho'}
\rho\equiv\Lambda_{IR}=\Lambda_{UV}\,e^{-\frac{\pi}{|\Delta_I|}}.
\ea
This relation is equivalent to~(\ref{Landau}), which has been found for the particular class of models studied in Section II. We thus argue that eq.~(\ref{rho'}) (or equivalently~(\ref{Landau})) characterizes a loss of conformality as a function of $\Delta$ (or, equivalently $x$) in a large class of models. The essential singularity at $\Delta=\Delta_c$ unambiguously associates the loss of conformality described here to a CPT.

We should compare~(\ref{rho'}) with~(\ref{rho}), where the phase transition was not understood as a function of $x$, but rather of the \textit{temperature}. The different factor of $\pi$ in the exponent is due to the different path followed to reach the critical point ($\Delta\rightarrow\Delta_c$ along the real axes in~(\ref{rho}), and $\Delta\rightarrow\Delta_c$ along the imaginary axes in~(\ref{rho'})), and may be associated to the real or imaginary nature of $\sqrt{D}$ in eq.(\ref{mass}).

\subsection{Modeling the strongly coupled branch}

The authors of~\cite{CL} looked for a weakly coupled description of the strong phase $\lambda>\lambda_{IR}$. Unfortunately, the examples they found do not manifest the desired non-trivial UV fixed point. Armed with the results of the previous Sections we propose a new candidate which, at least in a leading large $N$ expansion, has the required properties.

We are interested in studying the effect of a departure from the conformal point $\lambda=\lambda_{IR}$. This departure is understood to be induced by the strong dynamics as soon as $\lambda>\lambda_{IR}$, but we certainly do not know what is the complete set of (UV) relevant deformations that correctly describes the strong branch.

We decide to approach the problem using an effective perspective. We collectively parametrize the deformations with a coupling $\bar f$, and employ a somewhat unconventional approach to perturbation theory in which we identify the unperturbed theory ($\bar f=0$) with the $SU(N)$ gauge theory defined at the point $\lambda=\lambda_{IR}$. The formal description of this dynamics is encoded in the path integral
\ba\label{pert}
\langle e^{i\int{\cal L}_{def}}\rangle_{CFT},
\ea
where ${\cal L}_{def}$ contains the $O(\bar f)$ perturbations. The unperturbed theory is defined in terms of the Green's function of the $SU(N)$ gauge theory with the condition $\lambda=\lambda_{IR}$. Perturbation theory in the collective coupling $\bar f$ is unconventional in the sense that the \textit{free} theory possesses connected $3,4,\dots$-point functions.

The theory~(\ref{pert}) may be formally used to describe the portion of the beta function above the horizontal axes in~FIG.~\ref{fig2} if we think of the couplings $\bar f$ as unknown functions $\bar f(\lambda)$ of the 't Hooft coupling. Let us scrutinize further our approach. 

Under our hypothesis, the deformation ${\cal L}_{def}$ should be irrelevant around the (IR) fixed point $\bar f=0$, and should be such that the theory~(\ref{pert}) describes a flow towards a non-trivial fixed point in the UV, i.e. that the deformed theory is asymptotically safe. Except for these two requirements, the nature of ${\cal L}_{def}$ is completely arbitrary, though. In other words, there is generally more than one possible description of the strong phase. The reason stems from the fact that we are attempting to define a theory at an IR fixed point, and as such it inevitably represents an effective description, very much like the chiral lagrangian or a $\varphi^4$ theory in $d=4$. A sensible theory, of which an unambiguous continuum limit exists, must be defined at an UV fixed point. We therefore see that our formalism~(\ref{pert}) enables us to describe an entire class of theories whose elements are characterized by inequivalent UV completions. With this observation in mind we follow~\cite{CL} and call this class QCD*.

A class of models describing a flow between an UV and an IR fixed points has been studied in detail in Section II. Let us therefore assume that ${\cal L}_{def}$ can be parametrized by
\ba\label{pert1}
{\cal L}_{def}=f(q_R^\dagger q_L)(q_L^\dagger q_R).
\ea
In the above expression ${\cal O}=q_R^\dagger q_L$ is a color singlet scalar transforming as a bifundamental of the flavor group; ${\cal L}_{pert}$ is a singlet of the chiral group. The resulting dynamics~(\ref{pert}) has been analyzed in the planar approximation, and it is known to develop an UV fixed point. Recalling the identification $\bar f=\bar f(\lambda)$ we are tempted to associate the UV fixed point $\bar f_{UV}$ to the non-trivial fixed point $\lambda_{UV}$ of the strong branch as dictated by $\bar f(\lambda_{UV})=\bar f_{UV}$. In this sense, the beta function of $\bar f(\lambda)$ correctly reproduces the strong phase in~FIG.~\ref{fig2}, as anticipated. Furthermore, the relation $\Delta_-+\Delta_+=4$ -- where $\Delta_\pm$ are the dimensions of the quark bilinear at the fixed points -- is satisfied. At the critical point $x=x_c$ the quark bilinear has dimension $\Delta_c=2$, and the quantum mechanical running of the coupling $\bar f$ is responsible for the CPT scaling~(\ref{rho}). In this limit the beta-function~(\ref{beta1}) becomes IR unstable in the region $\lambda>\lambda_c$ signaling chiral symmetry breaking $\langle{\cal O}\rangle\neq0$. This motivates our choice ${\cal O}=q_R^\dagger q_L$. The emerging physics agrees with the results of~\cite{Gies}.

Notice that, a posteriori, the choice~(\ref{pert1}) is justified: for $\lambda_{IR}$ slightly below $\lambda_c$ the scalar operator ${\cal O}=q_R^\dagger q_L$ has scaling dimension close to $2$, and it seems reasonable to expect that the leading (= most relevant) irrelevant deformation appearing in~(\ref{pert}) should be parametrized by~(\ref{pert1})~\footnote{As emphasized above, this is an arbitrary choice of deformation defining a specific element of QCD*. Lattice simulations should tell us if additional operators are generated by the QCD dynamics, and ultimately which element of QCD* correctly describes the strong branch, see also~\cite{Pallante}.}.

We will not be able to prove whether the fixed point persists at next to leading order -- and therefore if~(\ref{pert}) plus~(\ref{pert1}) defines an element of QCD* even after the inclusion of $1/N$ corrections -- but we will give a suggestive argument in favor of the $1/N$ renormalizability of the theory (which would eventually imply the existence of the UV fixed point!). Our perturbation can be equivalently written in terms of an auxiliary field as
\ba\label{QCD*}
{\cal L}_{def}=q_R^\dagger q_L\sigma^\dagger+h.c.-\frac{1}{f}\sigma^\dagger\sigma,
\ea
where the scalar $\sigma$ transforms as $\sim(N_f,\bar N_f)$ of the flavor $SU(N_f)\times SU(N_f)$. In terms of the new variables, the only \textit{perturbative coupling} appears to be the dimensionless quantity in front of the Yukawa operator $\bar qq\sigma$. The usual proof of the renormalizability of a given theory relies on the identification of the primitive divergent graphs and their associated counterterms. The bottom line is that a theory is said to be renormalizable provided all relevant and marginal operators allowed by the symmetries are included in the formulation, and provided no irrelevant deformation -- which could potentially spoil the smooth UV limit -- is present. This conclusion is clearly in agreement with our understanding of the Wilsonian RG flow, and it will be taken as our formal definition of renormalizability. We would like to show that, at leading order in $1/N$ and for $\Delta$ lying in the range~(\ref{range}) with $\Delta_c=2$, the lagrangian~(\ref{QCD*}) passes this non-trivial check.

In fact, it is easy to see that the field $\sigma$ has scaling dimension $d-\Delta$, and hence that its mass term is a relevant operator for $\Delta\geq\frac{d}{2}$. Similarly, the canonical dimension of its kinetic term is $2(d-\Delta)+2$ and it is therefore irrelevant if $\Delta<\frac{d+2}{2}$. If the latter bound is satisfied then the scalar $\sigma$ would remain auxiliary at the quantum level, and would guarantee the quantum mechanical equivalence between~(\ref{pert1}) and~(\ref{QCD*}). Consider then its potential interactions. Any odd power is forbidden by the unbroken flavor symmetry and we should limit our discussion to even powers. It is sufficient to focus on the most relevant, $\sigma^4$. Again, this operator is irrelevant if $4(d-\Delta)>d$, namely if $\Delta<\frac{3d}{4}$. 

The conclusion is that the lagrangian~(\ref{QCD*}) contains all the relevant operators allowed by the symmetries -- at least at leading order in $1/N$ -- provided $\Delta<\frac{3d}{4}$ and $\Delta<\frac{d+2}{2}$, and $\Delta\geq\frac{d}{2}$. In $d=4$ this is the range~(\ref{range}) -- with $\Delta_c=2$ -- in which we expect to find $\Delta$ when the strong dynamics (the underformed CFT) is in the conformal window. When subleading $1/N$ corrections are taken into account the scaling dimensions do not algebraically sum and our argument fails. As far as $N$ is large, however, the corrections are small and the scaling laws do not drastically change. This is basically what happens in ordinary perturbation theory and may be an indication of a well defined continuum limit for the theory~(\ref{pert}) plus~(\ref{pert1}) beyond the leading $1/N$ order.

We can verify our scaling argument for the trivial example $\lambda_{IR}=0$, i.e. the NJL model. Given the classical value $\Delta=d-1$, in this case we precisely find the expected condition $d<4$~\cite{Parisi}. At $d=4$ the scalar $\sigma$ becomes dynamical and the equivalence between~(\ref{pert1}) and~(\ref{QCD*}) is no more satisfied if $\lambda_{IR}=0$. In this latter case we should include $\sigma$ as a new degree of freedom in the lagrangian, thus obtaining a Yukawa theory of the form considered in~\cite{CL}.

\section{On the unparticle physics}




Howard Georgi recently pointed out that an IR conformal dynamics coupled to the standard model can lead to peculiar phenomenological signatures. This framework goes under the name of unparticle physics~\cite{Georgi}.

The effect of the electro-weak (EW) vacuum on the unparticle physics has been extensively studied in the literature, starting with~\cite{Fox}. The generic implications of a coupling of the form $H^\dagger H{\cal O}$ -- with ${\cal O}$ being a CFT operator and $H$ the standard model Higgs doublet -- are a breaking of conformality below the EW symmetry breaking scale and the emergence of a mass gap in the unparticle sector (see~\cite{Piai} for a non-generic scenario). At the scale of the mass gap the resulting theory is very model-dependent~\cite{Strassler1}, and an effective study of the phenomenological signatures becomes difficult. Our study of double-trace deformations offers the opportunity to discuss the physics of a class of models in which the effect of the Higgs vacuum expectation value on the CFT is calculable within the large $N$ framework.

Assume that the unparticle operator ${\cal O}$ is a singlet under the standard model gauge group but carries some unbroken flavor symmetries, in analogy with the models discussed before. This possibility has been addressed in~\cite{Delgado}. Then the leading interaction between the CFT and the standard model is parametrized by the coupling
\ba\label{ch}
-\frac{c_h}{{\cal M}^{2\Delta-2}}H^\dagger H{\cal O}^2,
\ea
where $c_h=O(1)$ and ${\cal M}$ is a large mass. The scaling dimension $\Delta$ of the operator ${\cal O}$ is assumed to satisfy the unitarity bound $\Delta>1$. The irrelevant nature of the coupling~(\ref{ch}) is expected not to drastically modify the IR physics, in particular the Higgs condensation. If this is indeed the case, the Higgs vacuum would induce a quadratic deformation of the CFT of the form we have studied in this paper, with $f=c_hv^2{\cal M}^{2-2\Delta}$. As a consequence, the deformed CFT would be characterized by the scale
\ba
|p_\sigma|^2={\cal M}^2\left(\frac{v}{{\cal M}}\right)^{\frac{2}{2-\Delta}}.
\ea

If $1<\Delta<2$ (stability requires $c_h>0$) we generally have a natural suppression of the resonant scale compared to the cutoff, $|p_\sigma|^2\ll {\cal M}^2$. In this case the propagator $\langle{\cal O}{\cal O}\rangle$ would respect the CFT predictions both in the IR -- namely, for $p^2\ll |p_\sigma|^2$ -- and the UV -- that is for scales $|p_\sigma|^2\ll p^2\ll{\cal M}^2$, manifesting a resonant behavior at scales $p^2\sim|p_\sigma|^2$. The resonance width is controlled by $\Delta$, and for $\Delta\sim2$ can be very broad.

For $\Delta>2$ (stability now requires $c_h<0$) the deformation $f{\cal O}^2$ is irrelevant and no interesting effect emerges since $|p_\sigma|^2\gg {\cal M}^2$. In this latter case the CFT would appear unaffected by the EW vacuum.

In a generic framework, operators of the form $-c{\cal M}^{4-2\Delta}{\cal O}^2$ are unavoidable. In the presence of that term the resulting dynamics would be calculable within the large $N$ expansion if $c>0$, and would lead to a new IR CFT in which the operator ${\cal O}$ acquires an IR dimension $\Delta>2$. In the presence of $c\neq0$, the coupling $c_h$ has clearly no dramatic effect on the CFT.

It is interesting to consider the case in which ${\cal O}$ is charged under the flavor symmetry of the standard model. In particular, if the minimal flavor violation prescription is at work (the only source of flavor violation comes from the Yukawa interactions), then, at leading order in the flavor violating terms, the dominant interactions of the CFT operator to the SM fermions $\psi$ include operators of the form~\footnote{A similar situation would arise if both the CFT operator and the Higgs are odd under a $Z_2$ symmetry.}
\ba\label{cf}
\frac{{c}_{ijkl}}{{\cal M}^{\Delta}}\bar\psi_i H\psi_j{\cal O}_{kl},
\ea
with $c_{ijkl}$ a generic matrix with $O(1)$ elements. At next to leading order in the flavor violating parameters one also expects an operator $H^\dagger H{\cal O}$ to appear; however, its coefficient would be loop suppressed and may be neglected in a first approximation.

The phenomenology induced by the operators~(\ref{ch}) and~(\ref{cf}) contains new features not considered in the unparticle literature so far. In addition to the decay processes $h\rightarrow{\cal O}{\cal O}$ and $\psi\rightarrow\psi{\cal O}$ (the former would potentially account for the elusive nature of the Higgs boson), we encounter a resonant phenomena in the channel $2\psi\rightarrow2\psi$ which would dominate over the contact terms discussed in~\cite{GI}. Even more striking would be the meson-unparticle mixing effects induced by~(\ref{cf}). Quantitatively speaking, the operator~(\ref{cf}) contributes to the chiral lagrangian below $\Lambda_\chi\sim200$ MeV a fermion mass term 
\ba
\sim\frac{v\Lambda_\chi^3}{{\cal M}^\Delta}Tr\left(U{\cal O}\right)+h.c.,
\ea
which would eventually contribute to the pion mass matrix, and in particular to the $\pi$-$K$ mixing. A conservative estimate tells us that the small parameter controlling the mixing is $\Lambda_\chi^3/v{\cal M}^2$, and that the corrections induced by the mixing are well within the experimental uncertainty for a cutoff in the TeV range.

\section{Discussion}

The universal dynamical content of~(\ref{theory}) has been revealed in the planar approximation. Of particular interest is the existence of a regime describing an RG flow between two fixed points of the renormalization group and manifesting a resonant behavior. We will see in~\cite{LV} that the resonance is mapped into an IR non-normalizable mode (Gamow state) on the gravity dual.

We studied in some details models with 4-fermion contact terms and speculated on the large $N$ renormalizability of the 4-fermion operator in the presence of non-abelian gauge interactions -- the formal description of the theory is given by~(\ref{pert})(\ref{pert1}).

The conjectured strongly coupled branch of non-supersymmetric 4-dimensional non-abelian gauge theories at zero temperature~\cite{CL} has been discussed. 
We argued that the existence of the strong branch is intimately connected with the emergence of a conformal phase transition, and ultimately with the existence of a critical dimension $\Delta_c>1$ for the quark condensate~\footnote{In order for the quark bilinear to break the chiral symmetry, it is crucial that $\bar\psi\psi$ has interactions, i.e. $\Delta_c>1$. In super QCD the would-be order parameter is a free field at the critical point, i.e. $\Delta_c=1$, and no chiral symmetry, nor CPT occur.}. A verification of these speculations can only be attained with a lattice simulation, and requires the identification of a conformal phase transition characterized by the order parameter $\rho\sim\Lambda_\chi$ given in~(\ref{rho}) or~(\ref{rho'}), depending on the path $\Delta\rightarrow\Delta_c$. 

Following our quenched QED example, we proposed a model for the strong branch in which the most relevant CFT deformation is parametrized by a 4-fermion contact term~\cite{CL}. The resulting dynamics predicts $\Delta_c=2$, in agreement with the independent arguments given in~\cite{CG}. 

Another potential consequence of the conjecture~\cite{CL} is the occurrence of a \textit{physical} UV fixed point in the IR-free phase $x>x_{af}$, which would render non-abelian gauge theories a highly non-trivial example of asymptotically safe scenario~\cite{Weinberg}.

In the last Section we briefly discussed the implications of our results on the unparticle physics. We showed that in a large class of IR conformal dynamics the effect of the electro-weak vacuum can be analyzed in the large $N$ approximation, and sketched a few phenomenological signatures of these models.

\acknowledgments
The author would like to thank Tanmoy Bhattacharya for interesting conversations, Stefano Cremonesi for useful comments, and especially Michael L. Graesser for numerous and helpful discussions. This work has been supported by the U.S. Department of Energy at Los Alamos National Laboratory under Contract No. DE-AC52-06NA25396.


 \end{document}